# Market Liquidity and Convexity of Order-Book

# ——Evidence From China

*Kenan Qiao*

Key Laboratory of Management, Decision and Information Systems, Chinese Academy of Sciences

E-mail: qiaokenan10@mails.ucas.ac.cn

**Abstract**: Market liquidity plays a vital role in the field of market micro-structure, because it is the vigor of the financial market. This paper uses a variable called convexity to measure the potential liquidity provided by order-book. Based on the high-frequency data of each stock included in the SSE (Shanghai Stock Exchange) 50 Index for the year 2011, we report several statistical properties of convexity and analyze the association between convexity and some other important variables (bid/ask-depth, spread, volatility, return.)

## Introduction

Liquidity is a fundamental factor in financial market. If a market lacks of liquidity, the market order can not be executed at a stable price level. Additionally, lack of liquidity also causes a considerable extra cost, bid-ask spread, for immediate transaction, and a higher volatility of return resulted by a more sensitive price impact. However, liquidity is difficult to be measured. Amount of variables have been used to detect the liquidity[1][2][3], such as the depth, spread and immediacy.

The best bid/ask-depth or bid-ask spread is widely used to measure the liquidity close to the market price. However, there are few micro-structure studies have found that those limit orders, beyond the best quote, also contain many useful liquidity information which would effects the price discovery process[4]. In another words, the whole order-book, included those limit orders in different quotes, could improve the transparency of market and reduce price volatility. Furthermore, limit orders beyond the best quote also provide us a prefect explanation for some interesting phenomena. Many scholars have studied the shape of price impact function and show us various results of the curve in different markets[5][6]. Nonlinearity[8] is a very significant feature of the curve and it can be explained perfectly by the shape of order-book[9][15][16].

Therefore, it is necessary to research a whole order-book for seeking potential liquidity information hiding behind those limit orders beyond best quote. However, recording or analyzing a high-frequency dataset of order-book is really a hard work. An unbroken order-book often contains 6 to 10 quotes and must record all the depth in those quotes. So a high-frequency dataset of order-book must be a mass data. Fortunately, with the development of electronic trading and computer technology, we are able to obtain the dataset now and many scholars have focused on this field. No surprisingly, their results are also not consistent[5][6][7], just like the study of price impact function. Maybe the heterogeneity of different countries and markets is the root cause of this diversity. This paper will use a high-frequency dataset covering all stocks in SSE 50 index in 2011 to research this issue and dig out the pattern in Chinese stock market.

## Market, Data and Measure of Order-Book

SSE is purely driven by an order-book and no market markers. A whole order-book always contains many quotes and corresponding depth in those quotes. From 'TinySoft.Net', we can obtain an order-book with 5 best quotes at both buy side or sell side, donated by $q_5^{bid}, q_4^{bid},...,q_1^{bid}, q_1^{ask}, q_2^{ask},...,q_5^{ask}$. The depth will be present by $d_5^{bid}, d_4^{bid},...,d_1^{bid}, d_1^{ask}, d_2^{ask},...,d_5^{ask}$ according to the same order. Bid means the quote and depth in buy side, and ask donates the quotes and depth in sell side. A traditional way to measure the sheet is calculate the spreads of adjoining quotes and the best depth, $d_1^{bid}$ or $d_1^{ask}$. However, High-Dimension is an intractable problem. For each of the order-book, we have two 10-demension vectors. Simply ignore the data beyond the best quota is not a good method, which will loss some useful information behind those quote. A new idea can provide us a possibility to combine all the data for building a liquidity variable. A possible order-book is exhibited here.

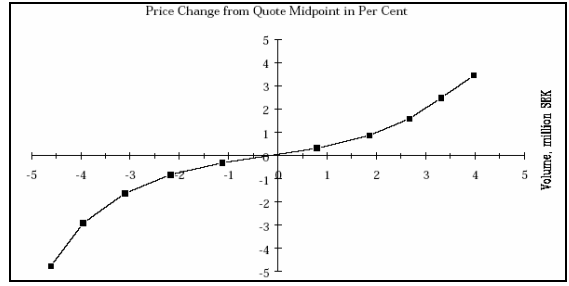

*Fig 1*

The y-axis is price pre-change from mid-quote,

$$w_i^{bid/ask} = \log q_i^{bid/ask} - \log mid, \text{ where } mid = \frac{q_1^{bid} + q_1^{ask}}{2}.$$

X-axis donates the accumulative depth, $D_i^{bid/ask} = \sum_{\alpha=1}^{i} d_\alpha^{bid/ask}$. In this case, we draw a convex curve which is steeper away from mid. However, it is not a universal discover. (Niemeyer, J., & Sandas, P. 1993) finds this curve at Stockholm Stock Exchange and the shape is in accordance with the theoretical model of (Glosten, L., 1994). But (Biais, B., Hillion, P., & Spatt, C. 1995) obtains a different conclusion, weakly concave, in Paris Bourse. In addition, a linear curve is also possible[10]. We find the convexity is various across the markets and countries. So researching the curve in Chinese market is really an interesting problem.

Now we need an index to measure the convexity of the curve based on a high frequency data. For stock k at day d, we take 30 samples in the interval $[t, t+5\min]$, frequency is 10sec. The order-book in each side will be denoted by $\{OB_{k,d,t,\theta}^{bid/ask}\}_{\theta=1}^{30}$. Furthermore, price pre-change from mid-quote and accumulative depth are denoted by $\{w_{k,d,t,\theta}^{bid/ask}\}_{\theta=1}^{30}$ and $\{D_{k,d,t,\theta}^{bid/ask}\}_{\theta=1}^{30}$. (Weber, P., Rosenow, B. 2005 etc.) discovers a power function is a perfect fitting for the sheet and price impact in Island ECN, so we estimate $w = W_{k,d,t}^{bid/ask} \cdot D^{c_{k,d,t}^{bid/ask}}$ to fit the order-book in $[t, t+5\min]$. A naive idea is to solve the following problem,

$$(P): \underset{W_{k,d,t}^{bid/ask}, c_{k,d,t}^{bid/ask}}{\text{minimize}} \sum_{\theta=1}^{30} \{\log w_{k,d,t,\theta}^{bid/ask} - [\log W_{k,d,t}^{bid/ask} + c_{k,d,t}^{bid/ask} \cdot \log(D_{k,d,t}^{bid/ask})]\}^2$$

$$\text{subject to} \quad W_{k,d,t}^{bid/ask} \geq 0$$

$$c_{k,d,t}^{bid/ask} \geq 0.$$

Bid side and ask side respectively, we use OLS to estimate the equation for obtaining an approximate fitting curve $w = W_{k,d,t}^{bid/ask} \cdot D^{c_{k,d,t}^{bid/ask}}$. We find that the results of $W_{k,d,t}^{bid/ask}$ and $c_{k,d,t}^{bid/ask}$ all subject to the restriction for each 5min interval needed to be estimated. So result of OLS is a good estimation of (P).

We point out that $W_{k,d,t}^{bid/ask}$ and $c_{k,d,t}^{bid/ask}$ have a meaningful economic interpretation. Obviously,

$$W_{k,d,t}^{bid/ask} = w|_{D=1}, \rho_{k,d,t;1}^{bid/ask} \overset{\Delta}{=} \frac{1}{W_{k,d,t}^{bid/ask}} = \frac{1}{w|_{D=1}}.$$

Thus, the $\rho_{k,d,t;1}^{bid/ask}$ is a 'initial density' of order-book. On the other hand,

$$\log w = \log W_{k,d,t}^{bid/ask} + \log D^{c_{k,d,t}^{bid/ask}},$$

$$\frac{dw}{w} = c_{k,d,t}^{bid/ask} \cdot \frac{dD}{D},$$

$$c_{k,d,t}^{bid/ask} = \frac{dD/dw}{D/w}.$$

So $c_{k,d,t}^{bid/ask}$ is 'marginal density' divided by 'average density'. It well known that if the $c_{k,d,t}^{bid/ask}$ is larger than 1, the curve would be convex, just like the figure 1. This pattern illustrate the density in best quote is larger than those quotes beyond the best quote. Conversely, if the $c_{k,d,t}^{bid/ask}$ is less than 1, the curve would be concave and the density of in best quote is less than those quotes beyond best quote. Therefore, $c_{k,d,t}^{bid/ask}$ is an effective index to measure the convexity of order-book.

## Basic Statistical Properties

From TinySoft.Net, we obtain a high-frequency data-set used to estimate the convexity for all the stocks in SSE 50 Index. For each stock, we have 200 available trading days at least and a trading day will be divided to 48 5min-intervals. Each of the intervals correspond a convexity, $c_{k,d,t}^{bid/ask}$, respectively in bid side and ask side. We also calculate some other meaningful

variables in the interval for the following analysis. Thay are

$$W_{k,d,t}^{bid/ask} : t= 1, 2, 3, 4......48,$$

$$\rho_{k,d,t;1}^{bid/ask} : t= 1, 2, 3, 4......48,$$

$$mid_{k,d,t} = \frac{\sum_{\theta=1}^{30}[(q_{k,d,t,\theta;1}^{bid} + q_{k,d,t,\theta;1}^{ask})/2]}{30} : t= 1, 2, 3, 4......48,$$

$$r_{k,d,t} = \log\frac{mid_{k,d,t}}{mid_{k,d,t-1}}, : t= 2, 3, 4......48,$$

$$\vartheta_{k,d,t} = \sum_{\theta=2}^{30}(\log\frac{mid_{k,d,t,\theta}}{mid_{k,d,t,\theta-1}})^2 : t= 1, 2, 3, 4......48.$$

$D_k$ is the number of trading days of stock k.

$\vartheta_{k,d,t}$ is a variable used to estimate the volatility of return[11]. (Barndorff-Nielsen O E, Shephard N. 2002) prove the follow equation

$$\sum_{k=1}^{n}(\log\frac{p_{k\cdot\Delta t}}{p_{k\cdot\Delta t-\Delta t}})^2 \xrightarrow{p} \int_{[0,T]} \sigma_{k,d,t,\theta}dB_{k,d,t,\theta} : n \to +\infty, \Delta t = \frac{T}{n}$$

under a semi-martingale assumption of process

$$d\log p_t = a_t dt + \sigma_t dB_t : t \in [0,T].$$

Some basic statistical properties of $c_{k,d,t}^{bid/ask}$ are reported in table 1.

|  | Mean | Std. Dev. | Median | Minimum | Maximum | Probability(t stat) |
|---|---|---|---|---|---|---|
| $\log c^{bid}$ | -0.6137 | 0.4006 | -0.6638 | -3.2102 | 2.4299 | 0.0000 |
| $\log c^{ask}$ | -0.5722 | 0.3972 | -0.6200 | -3.1530 | 2.5239 | 0.0000 |
| $\log c^{bid} - \log c^{ask}$ | -0.0414 | 0.3632 | -0.0431 | -3.3603 | 4.1452 | 0.0000 |

*Table 1. Summary Statistics*

Both bid side and ask side, we discover a significant negative $\log c^{bid/ask}$ which are at odd with previous empirical result (Niemeyer, J., & Sandas, P. 1993) and theoretical result (Glosten, L., 1994). The finding implies more limit orders stake in quotes away the mid-quote. This pattern may caused by lower transparency of market and different trading mechanism, T+1 and no market marker. A lower transparency market will increase the cost of adverse selection, so uninformed trader would willing to submit a higher quote at the cost of an extra waiting. Additionally, T+1 thoroughly eliminate high-frequency traders to get benefit from bid-ask spread, so depth in best quote will decrease significantly.

## Time-Difference Correlation Analysis

Analyzing the time-difference correlation is a fundamental stage when we research the dynamic process of $c_{k,d,t}^{bid/ask}$. We will calculate the auto-correlation coefficient by following equation,

$$\upsilon_{\Delta t}^{bid/ask} = \frac{\sum_{k,d} coeff(c_{k,d,t}^{bid/ask}, c_{k,d,t-\Delta t}^{bid/ask})|_t}{\sum_{k=1}^{50} D_k}.$$

Must points out, $\upsilon_{\Delta t}^{bid/ask}$ is an average result of all stocks and days. The result is displayed in figure 2,3.

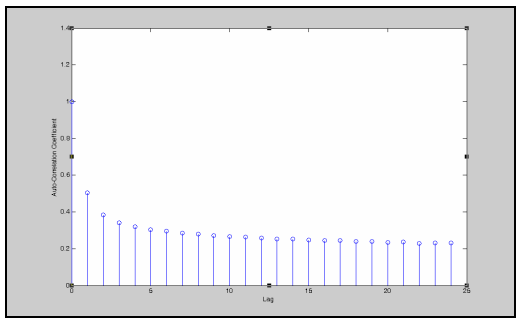
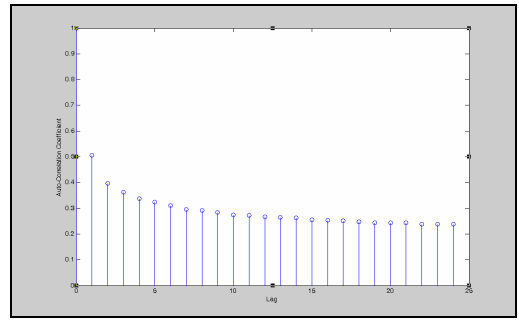

*Fig 2. auto-correlation coefficient(bid side)*  *Fig 3. auto-correlation coefficient(ask side)*

Obviously, there is a heavy tail in the figure. On there other hand, when we partial Coefficient by a similar way, we can find the coefficients decay rapidly. Therefore, AR model will be suitable to describe the dynamic process of $c_{k,d,t}^{bid/ask}$.

Using $\upsilon_{\Delta t}^{bid/ask} = a^{bid/ask} \cdot \Delta t^{-b^{bid/ask}}$ to fit the curve is an effective method. We sorely concern whether $b^{bid/ask}$ is less than 1, because it is an evidence of long-term memory[12]. Now we estimate the equation

$$\log \upsilon_{\Delta t}^{bid/ask} + \log \Delta t = \alpha - \beta \log \Delta t + \varepsilon_{\Delta t}^{bid/ask}, \alpha = \log a^{bid/ask}, \beta = b^{bid/ask} - 1.$$

The result is reported in table 2,3.

| Bid side | Estimation | Pvalue(t stat) | $R^2$ |
| --- | --- | --- | --- |
| $\alpha$ | -0.7965 | 0.0000 | —— |
| $\beta$ | -0.779 | 0.0000 | —— |
| Equation | —— | —— | 0.9977 |

Table 2. Regress on bid side

| Ask side | Estimation | Pvalue(t stat) | $R^2$ |
| --- | --- | --- | --- |
| $\alpha$ | -0.7578 | 0.0000 | —— |
| $\beta$ | -0.7781 | 0.0000 | —— |
| Equation | —— | —— | 0.9988 |

Table3. Regress on ask side

We find $b^{bid/ask}$ is significantly less than 1 on both sides, so $c_{k,d,t}^{bid/ask}$ may has long-term memory. However, a more reliable result must base on a more accurate statistic test. Here I point out there is an anti-correlation of $c_{k,d,t}^{bid/ask}$ short-term fluctuation. Donate

$$\kappa_{k,d,t}^{bid/ask} = \log c_{k,d,t}^{bid/ask} - \log c_{k,d,t-1}^{bid/ask}, t > 1.$$

The auto-correlation coefficient of $\kappa_{k,d,t}^{bid/ask}$ be calculated by

$$\upsilon_{\Delta t}^{bid/ask} = \frac{\sum_{k,d} coeff(\kappa_{k,d,t}^{bid/ask}, \kappa_{k,d,t-\Delta t}^{bid/ask})|_t}{\sum_{k=1}^{50} D_k}.$$

We report the result in figure 4,5.

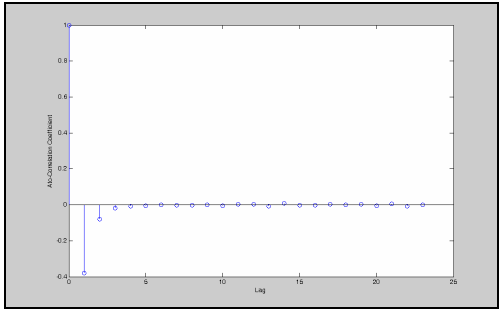 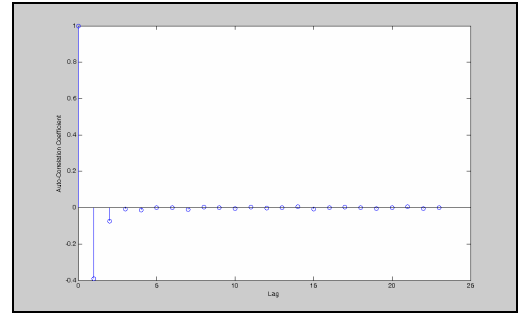

*Fig 4. auto-correlation coefficient(bid side)*    *Fig 5. auto-correlation coefficient(ask side)*

We can see that the coefficient is significant negative close to 0. A regress model can obtain a more reliable result. Estimate the equation

$$\kappa_{k,d,t}^{bid/ask} = a + b\kappa_{k,d,t-1}^{bid/ask} + \varepsilon\big|_{k,d,t}^{bid/ask}.$$

The result is reported in table 4.

| Bid side | Estimation | Pvalue(t stat) | $R^2$ |
| --- | --- | --- | --- |
| a | 0.0000 | 0.9973 | —— |
| b | -0.3790 | 0.0000 | —— |
| Equation | —— | —— | 0.1436 |

*Table 4. Regress on bid side*

| Ask side | Estimation | Pvalue(t stat) | $R^2$ |
| --- | --- | --- | --- |
| a | 0.0000 | 0.9961 | —— |
| b | -0.3896 | 0.0000 | —— |
| Equation | —— | —— | 0.1518 |

*Table 5. Regress on ask side*

Even thought the $R^2$ is less than 0.2, the coefficient, b, is significant on both side. This means the order-book has a reverse effect in short-term. Additionally, a is not significant on both side. So

$$\lim_{t\to+\infty}\kappa_{k,d,t}^{bid/ask} = \lim_{t\to+\infty}(\log c_{k,d,t}^{bid/ask} - \log c_{k,d,t-1}^{bid/ask}) = 0.$$

Therefore, there is an equilibrium point of $c_{k,d,t}^{bid/ask}$.

## Intraday Pattern

Numerous researches have found an intraday pattern of liquidity variable[13][14], spread and depth. Jain and Joh(1988) report a statistically significant U-shaped pattern in stock trading volume at the NYSE. Foster and Viswanathan (1993) test and reject the hypothesis of equal volumes across different hours of the trading day (using data from the NYSE and the AMEX). Those finding can be explain by asymmetric information and process of price discovery.

Here I will display the intraday pattern of convexity $c_{k,d,t}^{bid/ask}$ which exhibits a monotonously increasing curve during a day. We donate

$$c_t^{bid/ask} = \frac{\sum_{k,d} c_{k,d,t}^{bid/ask} / \tau_{k,d}^{bid/ask}}{\sum_k D_k}, where\ \tau_{k,d}^{bid/ask} = \frac{\sum_{t=1}^{48} c_{k,d,t}^{bid/ask}}{48}.$$

To make different stocks and days comparable, we normalize the $c_{k,d,t}^{bid/ask}$ by $\tau_{k,d}^{bid/ask}$ and calculate a mean at time t. The result is reported in figure 6,7.

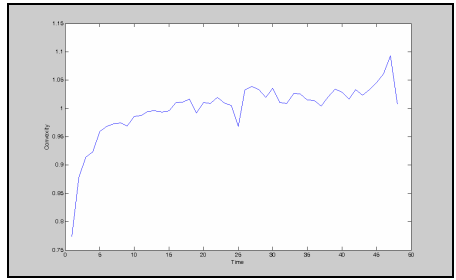 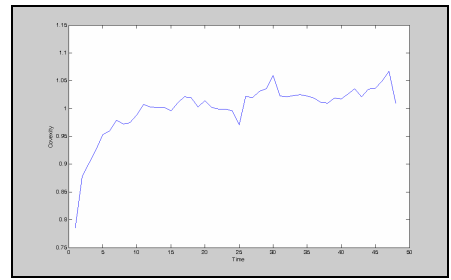

*Fig 6. intraday pattern of convexity(bid side)*   *Fig 7. intraday pattern of convexity(ask side)*

We find an upward intraday pattern of convexity. At the beginning of a trading day, traders willing to submit a higher quote to decrease the cost of adverse selection due to asymmetric information. As the transaction proceeds, information spreads in the market and cost of adverse selection decreasing rapidly. Asset price tents to an equilibrium point and traders submit more limit orders to best quote. Then the convexity increases significantly.

## Dynamic Adjustment of Convexity

Time-difference correlation analysis reveals the AR model is suitable for describing the dynamical process of convexity on both sides. However, we can involve more variables can impact the adjustment of convexity, such as return and volatility. We can find a significant coefficient in a regress model. This phenomenon implies some characteristics in financial market and can be explained by investor behavior.

For each stock, we estimate the following equation respectively,

$$\log c_{k,d,t}^{bid/ask} = \alpha^{bid/ask} + \beta^{bid/ask} \cdot \log c_{k,d,t-1}^{bid/ask} + \gamma^{bid/ask} \cdot (\log c_{k,d,t}^{bid/ask} - \log c_{k,d,t-1}^{bid/ask}) + \lambda^{bid/ask} \cdot r_{k,d,t} + \eta^{bid/ask} \cdot \vartheta_{k,d,t} + \varepsilon_{k,d,t}^{bid/ask}, t > 1.$$

Table 6,7 displays the result of our estimation of the equation.

| Bid side | Mean of K stocks | Number of coefficients significantly negative ( at [5%,10%] level) | Number of coefficients significantly positive ( at [5%,10%] level) | Mean r square |
|---|---|---|---|---|
| $\alpha^{bid}$ | -0.3311 | [50,50] | [0, 0] | —— |
| $\beta^{bid}$ | 0.4386 | [0, 0] | [50,50] | —— |
| $\gamma^{bid}$ | -0.114 | [50,50] | [0, 0] | —— |
| $\lambda^{bid}$ | 7.4308 | [0, 0] | [40, 46] | —— |
| $\eta^{bid}$ | -1647.9664 | [45,47] | [0, 0] | —— |
| Equation | —— | —— | —— | 0.1711 |

*Table 6. Regress on bid side*

| ask side | Mean of K stocks | Number of coefficients significantly negative ( at [5%,10%] level) | Number of coefficients significantly positive ( at [5%,10%] level) | Mean r square |
|---|---|---|---|---|
| $\alpha^{ask}$ | -0.2928 | [50,50] | [0, 0] | —— |
| $\beta^{ask}$ | 0.4557 | [0, 0] | [50,50] | —— |
| $\gamma^{ask}$ | -0.1341 | [50,50] | [0, 0] | —— |
| $\lambda^{ask}$ | -6.8807 | [47, 50] | [0, 0] | —— |
| $\eta^{ask}$ | -2012.1898 | [46, 46] | [0, 0] | —— |
| Equation | —— | —— | —— | 0.1701 |

*Table 7. Regress on ask side*

We can see that most of the stocks present a significant positive $\lambda^{bid}$ on bid side. This result reveals a positive return will impact the expectation of traders who will submit a higher quote for avoiding an extra cost in the future due to a higher expected price. On the contrary, a negative return will influent trades to submit a lower quote for an extra benefit. On the other hand, on ask side we obtain a negative $\lambda^{ask}$. This finding can be explained similarly with bid side. A positive return impel the traders submit a higher quote because of a higher expected return, and a negative return has an opposite effect.

Most of $\eta^{bid/ask}$ are significantly negative in our regress model and the economic interpretation of the phenomenon is very visual. A higher volatility reveals more serious asymmetric information which impel the traders submit a quote away from the mid-quote for avoiding a cost of adverse selection. Some similar result has been that depth and spread of an order-book also be impacted by volatility, but our finding focuses on the shape of the order-book.

## Contribution to Price Discovery

Following the analysis of (Weber, P., Rosenow, B. 2005), we consider an assuppositional return on order-book,

$$r_{k,d,t}^{book} = W_{k,d,t}^{ask} \cdot (V_{k,d,t}^{buy})^{c_{k,d,t}^{ask}} - W_{k,d,t}^{bid} \cdot (V_{k,d,t}^{sell})^{c_{k,d,t}^{bid}}.$$

$V_{k,d,t}^{buy/sell}$ is the initiative volume of buyer/seller. $r_{k,d,t}^{book}$ is the price impact by the volume trading with the order-book, because $W_{k,d,t}^{ask} \cdot (V_{k,d,t}^{buy})^{c_{k,d,t}^{ask}}$ and $W_{k,d,t}^{bid} \cdot (V_{k,d,t}^{sell})^{c_{k,d,t}^{bid}}$ is the return with a constant order-book. However, many limit orders arrive at order-book and be transacted by market orders. Those limit order can not be discovered by a time average order-book, because they are momentary. Therefore, $r_{k,d,t}^{book}$ usually doesn't equal to $r_{k,d,t}$. But we find $r_{k,d,t}^{book}$ concludes more information about price in the future which can not be explained by return. We consider the following equation.

$$\log p_{k,d,t} = \alpha + \beta \cdot \log p_{k,d,t-1} + \gamma \cdot r_{k,d,t-1}^{book} + \lambda \cdot r_{k,d,t-1} + \varepsilon_{k,d,t}.$$

Estimation of coefficient is reported in table 8.

|   | Mean of K stocks | Number of coefficients significantly negative ( at [5%,10%] level) | Number of coefficients significantly positive ( at [5%,10%] level) | Mean r square |
| --- | --- | --- | --- | --- |
| $\alpha$ | 0.0001 | [2, 3] | [3,3] | —— |
| $\beta$ | 0.9999 | [0, 0] | [50,50] | —— |
| $\gamma$ | 0.2970 | [0, 0] | [47, 49] | —— |
| $\lambda$ | 0.1511 | [0, 0] | [49,49 ] | —— |
| Equation | —— | —— | —— | 0.9088 |

*Table 8*

We find a positive effect of $r^{book}_{k,d,t-1}$, because $\gamma$ is significantly larger than 0 for most of our stocks. For bid side, a large convexity implies more depth stacked closing the best quote than lower and an initiative selling impact minor price changes. Thus the $r^{book}_{k,d,t-1}$ will be less than an order-book with a less convexity on bid side. According to the positive effect of $r^{book}_{k,d,t-1}$, $\log p_{k,d,t}$ will increase. That means a thick order-book close to mid-quote on bid side will increase the price in next 5min. On the contrary, implementing a similar analysis, a thick order-book close to mid-quote on ask side will decrease the price in next 5min. Therefore, the shape of order-book has a contribution to price discovery, even though the regress model contains observed return, $r_{k,d,t-1}$. By the way, $r_{k,d,t-1}$ also has an opposite effect due to a positive $\lambda$ and this illustrates Chasing behavior of traders.

## Conclusion

In this paper, we analyze a liquidity variable named convexity and report the summary statistics. We find a positive convexity which contradicts to previous conclusion. The result of time-difference correlation analysis implies a long-term memory of convexity and an anti-correlation of $c^{bid/ask}_{k,d,t}$ short-term fluctuation. Similar with traditional research of spread, we find intraday pattern of convexity which illustrates convexity will increase and approaches to an equilibrium during a day. This conclusion can be explained by information spreads. Additionally, we use a regress model estimate the dynamic adjustment of convexity and discover a significant impact of return and volatility on both side. Furthermore, we also find the order-book contributes to price discovery and has opposite effects to price in the future. Those finds illustrate convexity is an effective measure of market liquidity, and the depth away from mid-quote also has significant effect to improve the transparency of financial market.

## Reference：